\documentclass[useAMS,usenatbib]{mn2e}
\usepackage{times,aas_macros}
\usepackage{rotating}
\input{epsf}

\title[]
{Re-examining the XMM-Newton Spectrum of the Black Hole Candidate
XTE J1652-453}

\author[C.-Y. Chiang et al.]
{Chia-Ying Chiang$^{1}$\thanks{E-mail: cychiang@ast.cam.ac.uk}, R.C. Reis$^{2}$, D. J. Walton$^{1}$ and A. C. Fabian$^{1}$\\
$^1$ Institute of Astronomy, University of Cambridge, Madingley Road, Cambridge CB3 0HA \\
$^2$ Department of Astronomy, University of Michigan, 500 Church Street, Ann Arbor, MI 48109, USA}

\date{Accepted 2012 June 25.  Received 2012 June 14; in original form 2012 March 1}
\pagerange{\pageref{firstpage}--\pageref{lastpage}} \pubyear{2012}

\begin{document}

\topmargin = -0.5cm

\maketitle

\label{firstpage}

\begin{abstract}
The \emph{XMM-Newton} spectrum of the black hole candidate XTE J1652-453 shows a broad
and strong Fe K$\alpha$ emission line, generally believed to
originate from reflection of the inner accretion disc. These data have been analysed by \citet{Hiemstra11} using a variety of phenomenological models. We re-examine the
spectrum with a self-consistent relativistic reflection model. A
narrow absorption line near 7.2 keV may be present, which if real is
likely the Fe {\sevensize XXVI} absorption line arising from highly
ionised, rapidly outflowing disc wind. The blue shift of this
feature corresponds to a velocity of about 11100 km s$^{-1}$, which
is much larger than the typical values seen in stellar-mass black
holes. Given that we also find the source to have a low inclination
 ($i \la 32^{\circ}$; close to face-on), we
would therefore be seeing through the very base of outflow. This
could be a possible explanation for the unusually high velocity. We
use a reflection model combined with a relativistic convolution
kernel which allows for both prograde and retrograde black hole
spin, and treat the potential absorption feature with a physical
model for a photo-ionised plasma. In this manner, assuming the disc
is not truncated, we could only constrain the spin of the black hole
in XTE J1652-453 to be less than $\sim$ 0.5 $Jc/GM^{2}$ at the 90\%
confidence limit.

\end{abstract}


\section{Introduction}

Black hole binaries (BHBs) are known to exhibit a number of active
accretion states, each with different spectral and timing characteristics.
Amongst these, the most prominent are the low/hard state (LHS), dominated
by hard powerlaw-like emission which most likely arises due to Compton
scattering, and the thermal-dominated state (TDS), where the X-ray
spectrum displays the strong presence of thermal emission from an
optically thick accretion disc. Between these two well known states lie a
number of intermediate states, in which the Comptonised emission and the
thermal disc can contribute comparable amounts to the radiated flux (see
\citealt{HomanStates2001} and \citealt{Remillard06} for a full description
of spectral states in black hole binaries).

It is often found that, in addition to this X-ray continuum, the
spectra of accreting black holes -- both stellar mass ($M_{\rm BH}
\sim 10~M_{\odot}$) and supermassive ($M_{\rm BH} \ga
10^{5}~M_{\odot}$) black holes -- display the presence of reflection
features, arising through irradiation of the ``cold'' accretion disc
by the high energy Comptonised X-rays. Of these, the Fe K$\alpha$
line is usually the most prominent feature. Although this feature is
atomic in nature and hence intrinsically narrow, the observed iron
emission line is often found to be very broad due to a combination
of the Doppler effect and gravitational redshift close to the black
hole \citep{Fabian89}. The first observational detection of a broad
Fe K$\alpha$ line was found in the X-ray spectrum of the Type I AGN
MCG-6-30-15 \citep{Tanaka95}, and such features have since been seen
in a very large number of objects ranging from neutron stars
(\citealt{ bhattacharyya07,  cackett09, cackett10, disalvo09,
reisns}), stellar mass black holes (\citealt{miller07review, blum09,
reis09spin, Hiemstra11}) and AGNs (\citealt{Tanaka95,  FabZog09,
miniutti09spin, schmoll09,  3783p1, 3783p2}). Furthermore,
\citet{Walton12} highlight the similarity of the broadened iron
lines observed in black hole binaries and active galaxies, strongly
supporting a relativistic disc reflection origin. Due to the fact
that these lines are emitted from regions close to the central black
hole, their profile offers information of the innermost regions of
the accretion flow, and can be used to determine the inner radius of
the accretion disc from which black hole spin can be inferred, under
the assumption that this radius represents the innermost stable
circular orbit (ISCO; \citealt{Bardeenetal1972}). The other major
signature of reflection is the Compton hump, a broad emission
feature at $\sim$30\,keV which arises due to the relative interplay
of photoelectric absorption of low energy photons and Compton down
scattering of high energy photons within the reflecting medium. This
feature is also frequently observed in both stellar mass and
supermassive black holes (see $e.g.$
\citealt{Zdziarski99,Zdziarski02,Miniutti04,Cadolle07,Walton10,
Reis10lhs,Chiang11}).

Being largely atomic in nature, reflection features are independent
of black hole mass, providing the ideal method to measure spin in
both stellar mass black holes and their supermassive counterparts.
It is also possible to measure the inner radius of the disc through
direct study of the thermal disc emission, but detailed knowledge of
the black hole mass and its distance are required
\citep{Zhangetal19971655qpo,DavisDoneBlaes2006,mcclintock06}.
Recently, \citet{Steiner09} have shown that both the reflection
features and the thermal continuum give similar estimates for the
spin of the black hole binary XTE\,J1550-564, undertaking one of the
first major attempts at cross calibrating these two
methods.

In addition to these emission processes, if there is substantial intervening
material along our line of sight to the black hole, this will imprint
absorption features onto the observed X-ray spectrum. Complex absorption
features are observed relatively frequently in both active galactic nuclei
and X-ray binaries, although with somewhat varying properties. In the latter
case, the material usually appears to be highly ionised and outflowing,
suggesting an origin in some kind of wind launched from the inner regions of
the accretion flow; For an excellent example, see the \emph{Chandra} and
\emph{XMM-Newton} grating observations of the Galactic black hole transient
GRO J1655-40 presented by \citep{Miller06}.

XTE J1652-453 is a new X-ray transient discovered as part of the
\emph{RXTE} Galactic bulge scan performed in 2009
\citep{Markwardt09}. Later observations by the \emph{Swift} and
\emph{RXTE} indicate the source is likely a black hole candidate
\citep{Markwardt09b}. Simultaneous \emph{XMM-Newton} and \emph{RXTE}
observations were taken during the decay of the 2009 outburst, in
which the source was found to be in a hard-intermediate state, and
originally presented by \citet{Hiemstra11}. These data show clear
evidence for a strong and broad iron emission feature and, through
considering a variety of physically motivated -- albeit largely
phenomenologically modelled  -- interpretations, the authors
concluded that the accretion disk must extend as far in as $\sim4$
gravitational radii ($R_{\rm G} = GM/c^{2}$) from the central black
hole. Under the standard assumption that this is the ISCO, this
radius equates to a spin of $\sim0.5$, although \citet{Hiemstra11}
argue that the disc may in fact be truncated at a radius larger than
the ISCO, and that this should be considered as a lower limit to the
intrinsic spin of the system. A further curiosity unearthed by these
authors is that, when using the fully self-consistent reflection
code {\small REFLIONX} \citep{Ross05} to account for the line
profile, the best fit model showed the presence of a possible
ionised absorption feature at $\sim$7.2 keV, which although
considered briefly, was ultimately dismissed as a potential
deficiency in the model.

The purpose of the present work is to revisit the spin constraint
obtained for XTE J1652-453 with the advent of the {\sevensize REFBHB}
\citep{Ross07} reflection model, which is self-consistently
calculated for use with high temperature accretion discs present in
black hole binary systems, and to investigate whether accounting for
the absorption-like feature in a physical manner has any effect on
this result. The paper is structured as follows: section 2 briefly
outlines the data reduction, section 3 details our analysis of the
broadband spectrum, section 4 discusses the results obtained and
finally section 5 summarises our conclusions.

\begin{figure}
\centering
\leavevmode \epsfxsize=8.5cm \epsfbox{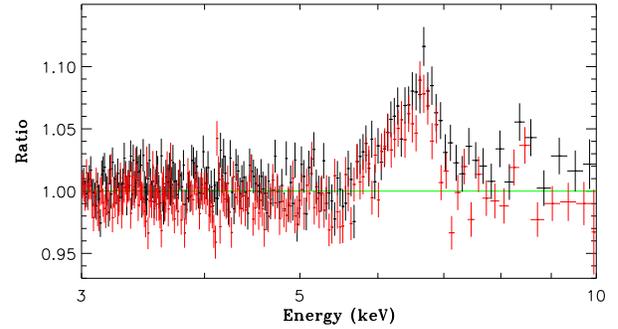}
\caption{The figure shows the broad Fe line in the \emph{XMM} PN
spectrum of XTE J1652-453. The black data points present the
spectrum without the CTI corrections, while the red ones stand for
the CTI-corrected spectrum. A mild gain shift between these
spectra can been seen in the figure. The line profiles in the
spectra are almost identical. The small peak around $\sim$ 8 keV
comes from the background.} \label{line}
\end{figure}

\section{Data Reduction}

We reduced the XMM-Newton data using the Science Analysis Software
(SAS) {\sevensize 10.0.1} with the latest calibration files. The
European Photon Imaging Camera (EPIC) PN was operated in timing
mode, and we extracted source spectrum of the PN data using the RAWX
columns in [30:46] following the standard procedures and the initial
work by \citet{Hiemstra11}.  As was shown by these authors, the
wings of the source point spread function (PSF) extend beyond the
CCD boundaries and therefore there are no areas in the chip that is
source free from which a background can be taken. However,
\citet{Hiemstra11} investigated in detail the effect of different
background models and concluded that this has little influence in
any of the fit parameters of interest. For this reason we follow
from their work and do not use a background and add the further note
that the source count rate is over two hundred times higher  than
that of  the  contaminated background.

As the EPIC-PN detector was operated in the fast-readout Timing
mode, the effects of charge-transfer inefficiency (CTI) and/or X-ray
loading (XRL) may not have been properly accounted for by the
standard reduction pipeline\footnote{see
http://xmm2.esac.esa.int/docs/documents/CAL-TN-0083.pdf}, resulting
in a mismatch between the energies of the instrumental edges
($\sim$2\,keV) in the data and in the response matrices. Although
strong residuals are not observed around these features in this
case, the previous work on this source attempted to correct for any
potential calibration uncertainties that remain with the {\sevensize
EPFAST} tool. However, as demonstrated by \cite{Walton12}, the
application of {\sevensize EPFAST} can result in an incorrect energy
scale at high energies, particularly around the iron line.
Therefore, we test briefly whether this tool has any strong effects
on the results obtained. We find that in this case, owing to the
much lower source count rate, {\sevensize EPFAST} does not apply any
significant correction, and the unmodified and {\sevensize
EPFAST}-modified data are fully consistent, as shown by Fig.
\ref{line}. However, owing to the presence of an unidentified excess
below ~1.5\,keV (such features features are frequently seen in
Timing mode observations$^{1}$) we only fit the PN data above 2.2
kev and use the The Reflection Grating Spectrometer (RGS) between
0.4--2.0 keV, similar to the approach taken by \citet{Hiemstra11}.
Similarly, we also used the RXTE Standard-2 data between 7 and 20
keV and the HEXTE data between 20 and 200 keV allowing for a
constant of normalisation between all different spectra.

\begin{figure}
\leavevmode \epsfxsize=8.5cm \epsfbox{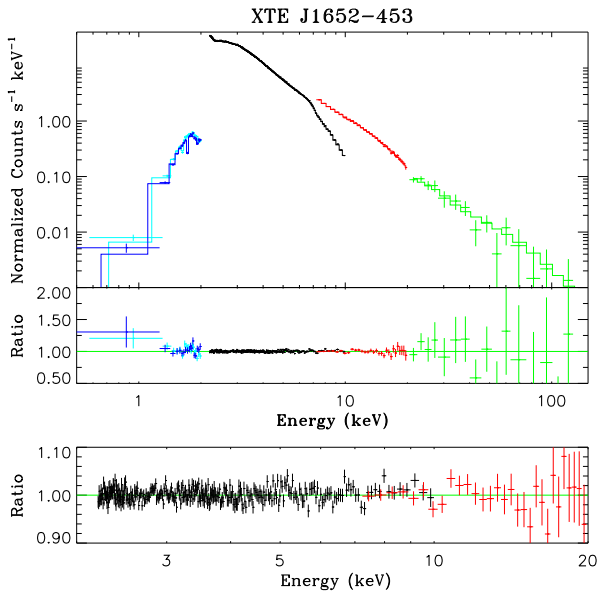}
\caption{The figure shows the best fit of the simultaneous
\emph{XMM-Newton} and \emph{RXTE} observations using model A. The
black data points in upper panel belong to the PN spectrum, while
the blue and cyan ones are of RGS1 and RGS2, the red and green ones
are of the \emph{PCA} and \emph{HEXTE} spectrum, respectively. The
lower panel the data/model ratio of the fitting. Data points are
re-binned for clarity.} \label{fitting_ref}
\end{figure}

\section{Data Analysis and Results} \label{section_da}

As mentioned in the introduction, the \emph{XMM-Newton} spectrum of
XTE J1652-453 shows a broad, skewed line profile (see Fig.
\ref{line}) that is likely  reflected emission  from the inner part
of the accretion disc around the black hole. We refer the reader to
\citet{Hiemstra11} for a detailed analysis of this source using a
variety of phenomenological models. To cross-check with their results, we begin by reproducing
the final model presented in that work, which consists of a powerlaw
continuum, a thermal disk component ({\sevensize DISKBB};
\citealt{diskbb}), and relativistic disc reflection. The latter
component is modelled with the {\sevensize REFLIONX} reflection code
\citep{Ross05}, with the relativistic effects provided by the
{\sevensize KERRCONV} \citep{kerrconv} convolution kernel. Galactic
absorption is also included, initially modelled with the {\sevensize
PHABS} photoelectric absorption code, with $N_{\mbox{\scriptsize
H}}$ free to vary, and we also include the additional absorption
components originally used by \citet{Hiemstra11}. We obtained a spin
parameter of $\sim 0.5 \pm 0.2$, in excellent agreement with the
published result, and the other parameters in the model also agree
well within statistical errors. The $\sim$ 7.2 keV absorption line
which has been reported by the authors is also present in our
spectrum. As previously mentioned, this feature was dismissed as a
potential deficiency in the model. In order to check the
validity of this claim, we proceed by modelling the broadband
continuum with a physically self-consistent model including the latest reflection and relativistic convolution
models.

\begin{figure}
\centering
\leavevmode \epsfxsize=8.5cm \epsfbox{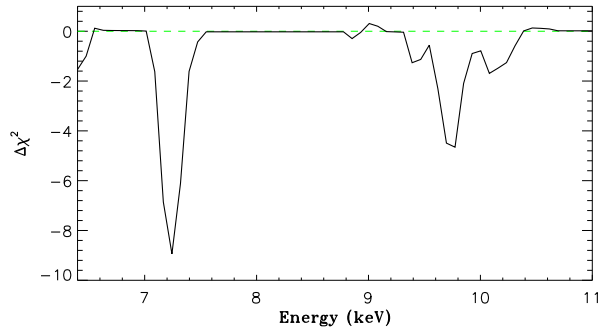}
\caption{The figure shows a search of energy space between 6.4 and
11 keV when fitting a narrow (10 eV) inverse Gaussian line to the
spectrum. The y-axis shows the amount of the $\chi^2$ changed after
putting the inverse line. } \label{contour_en}
\end{figure}

First we replaced the Galactic absorption model {\sevensize PHABS}
with the latest high resolution model {\sevensize TBNEW}\footnote{http://pulsar.sternwarte.uni-erlangen.de/wilms/research/tbabs/ }
(Wilms, Juett, Schulz, Nowak, 2012, in preparation), using the new solar abundances presented in that
work. We then replaced the {\sevensize REFLIONX} reflection model
with {\sevensize REFBHB} (\citealt{Ross07}), and the {\sevensize
KERRCONV} convolution kernel with {\sevensize RELCONV}\footnote{http://ww.sternwarte.uni-erlangen.de/~dauser/research/relline/index.html}
\citep{Dauser10}. The {\sevensize REFBHB} reflection model has been
calculated specifically for use with black hole binary systems,
allowing for a hot accretion disc as opposed to the cool disc
temperature with which {\sevensize REFLIONX} was calculated, and
including the thermal emission from this component self consistently
both in the spectrum and when determining the disc ionisation. The
key free parameters of {\sevensize REFBHB} are the photon index of
the illuminating continuum ($\Gamma$), which we require to be the
same as the powerlaw continuum, the blackbody temperature of the
accretion disc ($kT$), the surface hydrogen density of the accretion
disc ($n_{\rm H}$), and the ratio of the illuminating and the
blackbody fluxes ($F_{\rm ill}/F_{\rm th}$). The {\sevensize
RELCONV} convolution model is very similar to {\sevensize KERRCONV},
but this model also allows for a retrograde spin parameter. We make
use of this code assuming the limb darkening effects outlined by
\cite{Laor91}. Note that, owing to the lack of background
correction, we also include an unresolved Gaussian feature at
8.1\,keV to account for the well known instrumental Copper feature\footnote{http://xmm.vilspa.esa.es/docs/documents/CAL-TN-0068-0-1.ps.gz}.
With this new model (hereafter model A), we find that the extra
absorption component at $\sim$ 1.4 keV mentioned in \citet{Hiemstra11} is
not obvious in our fit, although the potential absorption features
at $\sim$7.2 and $\sim$10\,keV are still present. This model
otherwise provides a good fit to the data, as shown in Fig.
\ref{fitting_ref}, and the results obtained are presented in Table
\ref{fitting}. We also include the results obtained by
\citet{Hiemstra11} with the {\sevensize DISKBB}+{\sevensize
REFLIONX} combination for ease of comparison.

There are some differences between our results and those of
\citet{Hiemstra11}. We obtained a higher Galactic absorption column
$N_{\mbox{\scriptsize H}}$ and a harder photon index $\Gamma$,
although the former at least is likely to be due to the different
reflection model used in our work.
%We have used the abundances presented in \cite{Wilms00} in our
%modelling, which usually give higher hydrogen column densities than
%the default abundances of \citet{Anders89}.
We obtain a good
agreement between the disc temperatures, as the temperature obtained
with {\sevensize REFBHB} is the effective disc temperature, which is
known to be a factor of $\sim$1.7 \citep{colourcor} less than the
surface (or ``colour'') temperature obtained when using {\sevensize
DISKBB}. The values of the emissivity profile index obtained in our
models are also in close agreement. The spin constraint of $a^* =
0.13\pm0.29$ obtained here, the main focus of this work, is much
weaker than that originally presented by \citet{Hiemstra11} ($a^* =
0.45\pm0.02$).

\begin{table*}
 \caption{The Table below lists the fitting parameters from different
  model components and $\chi^{2}$ obtained by different models. The Galactic absorption column
$N_{\mbox{\scriptsize H}}$ is given $10^{22}$
 cm$^{-2}$. The index stands for the emissivity profile index in the
 convolution model ({\sevensize RELCONV} for model A-C and {\sevensize KERRCONV} for the work by
 \citet{Hiemstra11}), and $a^*$ and $i$ stand for the spin parameter and the inclination angle, respectively.
 The absorption column $N_{i\mbox{\scriptsize H}}$ of the absorbing component is also given in $10^{22}$
 cm$^{-2}$, while the ionisation parameter $\xi$ is in erg cm
 s$^{-1}$. $n_{\rm H}$, $kT$ and $F_{\rm ill}/F_{\rm th}$ are parameters of the {\sevensize REFBHB} component.
 $^{*}$A broken emissivity profile has been used in this model. The emissivity index has been set to be 0 for
 inner part of the accretion disc and 3 for the outer region. }
\label{fitting}
\begin{tabular}{@{}lcccc}
\hline\hline
parameter & Model A  & Model B & Model C & Hiemstra et al.\\
\hline\hline
$N_{\mbox{\scriptsize H}}$ ($10^{22}$ cm$^{-2}$) & $7.07^{+0.07}_{-0.10}$ & $7.02\pm0.11$ & $7.01\pm0.10$ & $6.73\pm0.01$\\
$\Gamma$ & $2.10^{+0.02}_{+0.05}$ & $2.09^{+0.03}_{-0.04}$ & $2.09^{+0.03}_{-0.04}$ & $2.16\pm0.02$\\
\hline
index & $2.6^{+0.0}_{-0.3}$ & $3.1^{+0.4}_{-0.3}$ & fixed* & $2.7\pm0.1$\\
$R_{\rm bk}$ ($R_{\rm G}$) & - & - & $9.1^{+4.0}_{-2.3}$ & -\\
$a^*$ & $0.13\pm0.29$ & $\la 0.6$ & unconstrained & $0.45\pm0.02$\\
$i$ & $\la 14$ $^{\circ}$ & $28\pm3$ ${^\circ}$ & $28^{+3}_{-2}$ ${^\circ}$ & $8.8\pm0.1$ $^{\circ}$\\
\hline
$N_{i\mbox{\scriptsize H}}$ ($10^{22}$ cm$^{-2}$) & - & $> 2.3 $ & $> 2.2$ & -\\
log $\xi$ & - & $> 3.94$ & $> 3.96$ & -\\
$z$ & - & $-0.036^{+0.008}_{-0.006}$ & $-0.036^{+0.008}_{-0.005}$ & -\\
\hline
$n_{\rm H}$ ($10^{19}$ cm$^{-2}$) & $5.2^{+0.3}_{-0.6}$ & $7.7^{+3.0}_{-2.2}$ & $7.6^{+1.5}_{-2.3}$ & -\\
$kT$ (keV) & $0.38\pm0.01$ & $0.38\pm0.01$ & $0.38\pm0.01$ & $0.59\pm0.01$\\
$F_{\rm ill}/F_{\rm th}$ & $0.6\pm0.1$ & $0.5\pm0.1$ & $0.5\pm0.1$ & - \\
\hline
$\chi^{2}_{\nu} ~(\chi^{2}/{\rm d.o.f.})$ & 1.04 (2177.6/2095) & 1.04 (2164.1/2092) & 1.04 (2164.5/2092) & 1.26 (460.2/365)\\
\hline\hline
\end{tabular}
\end{table*}

\subsection{Potential Absorption in the Fe-K Band}

The potential absorption features at $\sim$7.2 and $\sim$10\,keV
were both originally reported by \citet{Hiemstra11}. In order to
investigate the statistical significance of these features with our
new model, we added an inverse narrow (line width $= 10$\,eV)
Gaussian line, and systematically varied the line energy between 6.4
and 11 keV in 60 steps. In Fig. \ref{contour_en} we plot the
$\Delta\chi^{2}$ improvement gained with the addition of the narrow
line as a function of the line energy. The potential features at 7.2
and 9.7\,keV are clearly picked out, although their overall
detection significances are not particularly high. The feature at
7.2 keV has the strongest detection significance of the two, but we
estimate that this is only just greater than 2$\sigma$; the
equivalent width obtained is $EW = - 20^{+13}_{-15}$\,eV. If real, the
most likely association of this feature is highly ionised iron
absorption arising in some outflowing disc wind. The closest
transition is Fe {\sevensize XXVI} K$\alpha$ at 6.97\,keV, although
this would imply an outflow velocity of $\sim$11000 km s$^{-1}$,
much higher than typically observed for black hole binary disc winds
(see \emph{e.g.} \citealt{Miller06}). Unfortunately we were unable
to find a suitable atomic association for the feature at 9.7\,keV
that would imply a similar outflow velocity, so the nature and
velocity (and even the presence) of the outflow remain unconfirmed.
Furthermore, since this second feature does not effect the Fe K
band, which is the primary focus of this work, we will not consider
it further.

Regardless of the strict statistical significance of the potential
Fe {\sevensize XXVI} absorption line, we wish to investigate what
effect modelling this feature in a physical manner has on the spin
constraint obtained. We therefore generated an ionised absorption
model using the {\sevensize XSTAR} photoionisation code (v2.2.0).
The ionising continuum was assumed to have a powerlaw shape over the
0.1--20 keV energy range, and we adopted an ionising luminosity
(1-1000 Ryd) of $10^{38}$ erg s$^{-1}$, a typical value for bright
stellar-mass black holes (the distance to this source is not yet
well constrained). The temperature of the photoionised gas was fixed
at $10^{5}$ K, the turbulent velocity $v_{\mbox{\scriptsize turb}}$
at 150 km s$^{-1}$, and the iron abundance at the solar value. The
free parameters of the generated model are the photon index of the
ionising continuum, which was required to be the same as the
powerlaw continuum component, the absorbing column density ($N_{\rm
H}$), the ionisation parameter ($\log\xi$) and the outflow velocity
(in the form of the redshift of the gas, $z$). We also tested an
absorption model in which the assumed ionising continuum had a
thermal shape, but the results obtained with these two models were
in excellent agreement so we only present the former case.

\begin{figure}
\leavevmode \epsfxsize=8.5cm \epsfbox{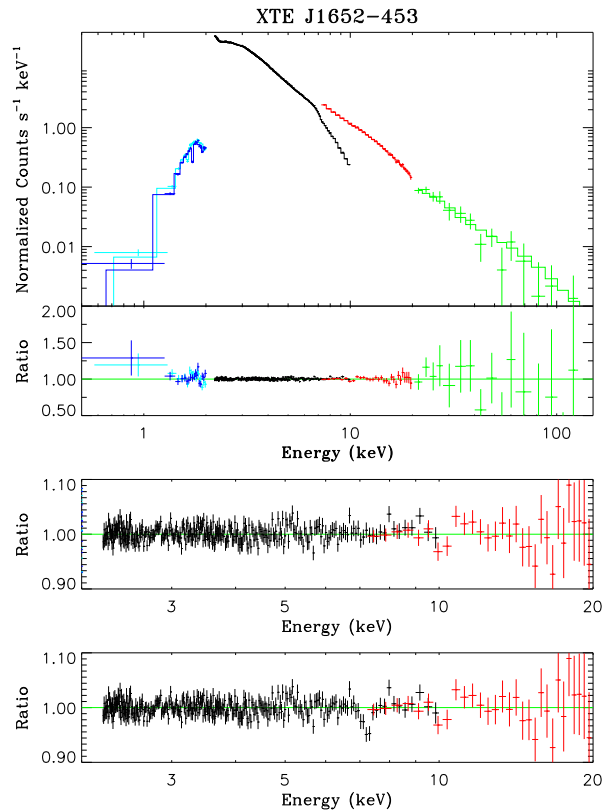}
\vspace{-0.7 cm} \caption{The upper part of the figure shows the
result fitted the data sets with model B, which includes an
absorbing zone. The middle part shows the zoom-in  data/model ratio
from 2 to 20 keV, while the lower part of the figure shows the
data/model ratio of the same energy range when the warm absorber is
taken out from model B. It can be clearly seen that the absorbing
zone only produces the absorption line around 7.2 keV.}
\label{fitting_warm}
\end{figure}

We added this absorption component to the model including
{\sevensize REFBHB} constructed above and re-fit the spectra (this
shall hereafter be referred to as model B). This model of course
fits the data well, and the results obtained are also quoted in
Table \ref{fitting}. The absorbing gas is highly ionised, in
accordance with the lack of absorption features in the RGS spectrum,
and, as demonstrated by Fig. \ref{fitting_warm}, only contritubes
the absorption line at 7.2\,keV.

\begin{figure}
\leavevmode \epsfxsize=8.5cm \epsfbox{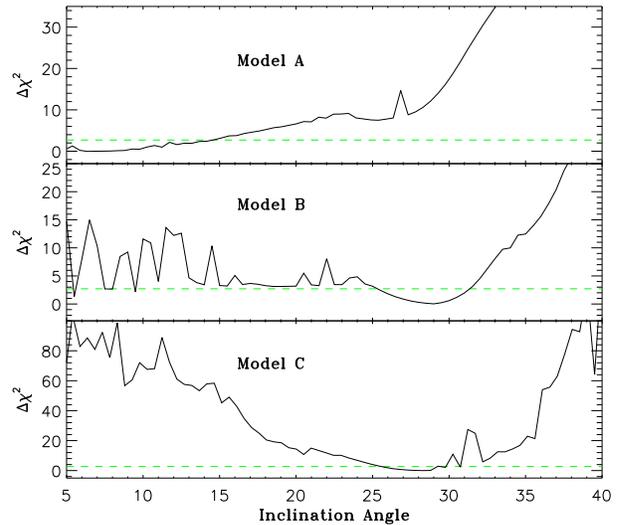}
\caption{Goodness-of-fit versus inclination angle of model A-C. For
model A an inclination angle $i \la 14^{\circ}$ is preferred. A
global minimum of approximately $28^{\circ}$ has been found with
model B. However, note a series of local minima located below
$10^{\circ}$. An inclination $>35^{\circ}$ is rejected at greater
than the 3$\sigma$ level.} \label{con_incl}
\end{figure}

Model A and B yield similar results in most of the fitting
parameters. However, when we include the photoionised absorber to
account for the possible Fe {\sevensize XXVI} feature at 7.2\,keV,
while there is a local minimum consistent with the result obtained
by \citet{Hiemstra11} and similar to the result obtained with model
A, we find that the globally preferred solution has a maximal
retrograde spin, with an upper limit of $a^* \la -0.4$. Fig.
\ref{con_spin} shows the 1-D $\chi^{2}$ confidence contours for the
spin parameter in both cases. A further difference is the value of
the inner disc inclination. The inclination obtained with the basic
reflection model, $i < 14^{\circ}$, is consistent with the
previously published result, but the inclusion of the ionised
absorber changes the inclination to $i = 28\pm3^{\circ}$, although
there are also a series of local minima at lower inclinations. Fig.
\ref{con_incl} shows the 1-D $\chi^{2}$ confidence contours for the
disc inclination for both cases.

These changes suggest there may be some degeneracies between the
values obtained for the spin and the inclination of the system.
Indeed, degeneracies in the blurring parameters have been found to
be a issue during the application of reflection models to a number
of other sources (see \emph{e.g.} \citealt{Nardini11}). In order to
fully investigate this possibiliy, we show the 2-D confidence
contours in the spin-inclination parameter space for model B in Fig.
\ref{con_a_incl}. The global minimum associated with a maximum
retrograde spin and indicated with a bold cross, is associated with
a higher inclination, while the local minimum at an intermediate
prograde spin is associated with a lower inclination. This would
appear to confirm that there is indeed a mild degeneracy between
these parameters. Given that it is not clear whether the feature at
7.2\,keV is real or simply a statistical fluctuation, the only limit
on the spin we can place with any confidence is $a^* \la 0.5$,
although a maximum regrograde spin remains a very intriguing
possibility.

\subsection{Large Scale Height Corona}

\begin{figure}
\leavevmode \epsfxsize=8.5cm \epsfbox{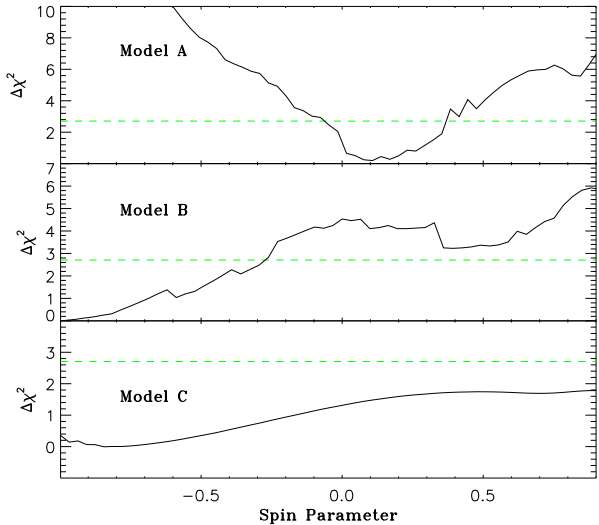} \caption{The
figure shows from top to down, the Goodness-of-fit versus spin
parameter for XTE J1652-453 with model A-C. The global minimum of
model B indicates a maximumly rotating retrograde black hole.
However, see degeneracies shown in Fig. 7. Model C does not give a
constrained value of the spin parameter.} \label{con_spin}
\end{figure}

An alternative geometry to a compact corona close to the
event horizon is the possibility that the corona is located at a
large vertical height from the accretion disc. In this scenario, the
emissivity index is expected to be flat up to a break radius
\citep{Vaughan04} where beyond this it reverts to the traditional
Newtonian value of 3. Such geometry, if not properly accounted for,
could potentially mimic a truncated accretion disc or a retrograde
spin. We test this possibility by setting the inner emissivity index
in model B to be 0 up to a break radius ($R_{\rm{bk}}$) and 3 beyond
that. This fit (model C, Table \ref{fitting}) yields a break radius
of $\sim 10 R_{\rm G}$, which suggests that the corona is located at
a similar vertical distance, and provides an equally good fit (reduced
$\chi^2$ = 2164.5/2092) to model B. From Fig. \ref{con_spin} we can
see that the spin parameter is unconstrained if the corona is
assumed to be at a large scale height.

\section{Discussion} \label{discussion}

\begin{figure}
\leavevmode \epsfxsize=8.5cm \epsfbox{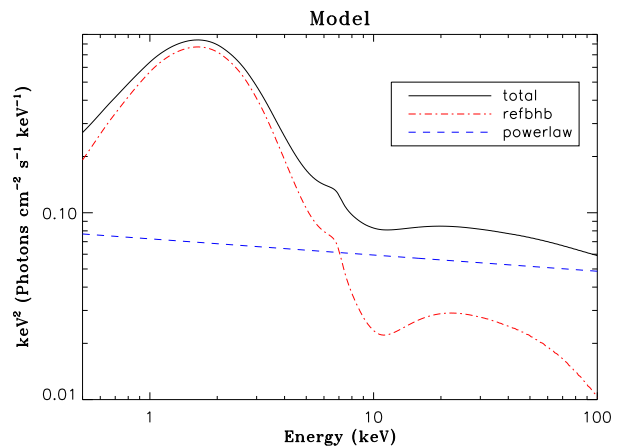} \caption{The figure
shows the main decomposed components of the model used in this work.
The total model is shown in black curve. Absorption components have
been removed from this figure.} \label{model}
\end{figure}

We have re-analysed the simultaneous \emph{XMM-Newton} and
\emph{RXTE} spectra of the black hole binary candidate
XTE\,J1652-453 obtained during the recent outburst displayed by this
source in 2009. Our work focuses in particular on the
relativistically broadened iron emission, with the purpose of making
use of the new {\sevensize REFBHB} reflection model to constrain the
spin of the black hole. The previous spin constraint of $a^* = 0.45
\pm 0.02$, published by \cite{Hiemstra11}, made use of the
{\sevensize REFLIONX} reflection model, which is calculated for use
primarily with AGN. We also make use of the new convolution kernel
{\sevensize RELCONV}, which accounts for the relativistic effects
present in the regions close to a black hole, and crucially allows
for black hole spins that are retrograde with respect to the
material orbiting in the accretion disc. The decomposed components of the model have been shown in Fig. \ref{model}.

We find first of all that, when using this new reflection model,
the constraint on the black hole spin obtained, $a^* = 0.13\pm0.29$,
is much looser than the previously published result. This most
likely reflects the differing scenarios for which {\sevensize
REFBHB} and {\sevensize REFLIONX} were created. Although the same
physics and atomic data are included in both models, the former
allows the disc temperature to be varied in a range suitable for the
hot discs around BHBs, while the latter assumes a cool disc
temperature of $kT = 10$\,eV, more suitable for the discs around
AGN. These higher temperatures naturally lead to a highly ionised
disc and a more significant contritution to the bredth of the
observed emission line from Compton scattering during the reflection
process, as Compton broadening becomes more important as the
temperature of the scattering electrons (and hence the disc
temperature) increases (\citealt{Pozdnyakov83}). Furthermore, we
also found that there is a mild degeneracy between the black hole
spin and the disc inclination, with higher inclinations broadly
leading to solutions with higher black hole spin. The combination of
these factors naturally make it more difficult to strongly constrain
the black hole spin, and lead to the updated constraints obtained in
this work being less restrictive.

\begin{figure}
\leavevmode \epsfxsize=8.5cm \epsfbox{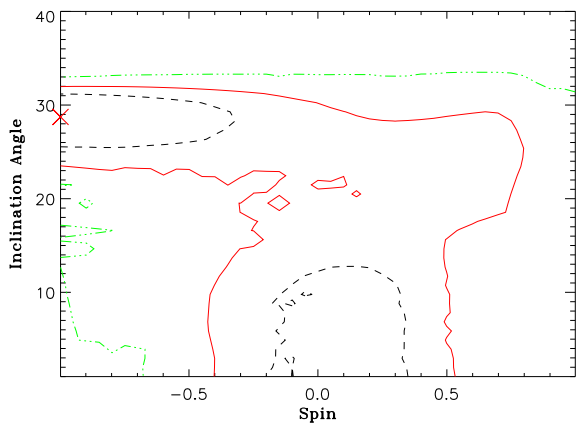}
\caption{The figure shows the contour plot of the spin parameter
against the inclination angle of model B. The 90\% contours span a
wide range, indicating the possible spin parameter could be any
value $\la$ 0.8. The red cross shows the best-fitting set of values
by model B.} \label{con_a_incl}
\end{figure}

Inspecting the residuals after modelling the spectrum with the
basic disc reflection interpretation, we confirmed the possible existence of a narrow
absorption feature at 7.2\,keV. This was originally highlighted by
\cite{Hiemstra11}, but dismissed as a deficiency in the model used in that work. The persistence of this potential
feature when using the best physically self-consistent black hole
binary reflection model currently available suggests that other
explanation should be pursued. We stress that the inclusion of a
Gaussian absorption feature only leads to a marginal improvement in
terms of the quality of fit, so we do not consider the detection of
this feature to be formally significant, and wish to reiterate the
strong possibility that it merely arises through statistical
fluctuations. However, it is also important to consider what effect
this potential feature might have on the spin constraint obtained.
With the inclusion of an ionised absorption
component, produced with the {\sevensize XSTAR} photoionisation
code, we find that the spin can only be constrained to be $a^* \la 0.6$ (when considering the 90\% confidence contour for the combination of spin and inclination). This difference most likely
arises due to the mild degeneracy between the spin and inclination
mentioned earlier (shown in Fig. \ref{con_a_incl}). The energy at which the residuals are seen is
roughly coincident with the blue wing of the emission line, which is
the region of the line profile that has the strongest influence on
the inclination obtained. The inclusion of this feature in the model
changes the inclination, and hence changes the spin constraint.

A black hole spinning in a retrograde manner with respect to the
orbital motion of its accretion disc is a very interesting and
unusual prospect. Retrograde spin can result in a large gap between the ISCO and the event horizon, in which
magnetic fields can gather and provide forces to power a jet. In fact there have been claims that retrograde
spin may launch stronger outflows than prograde spin (e.g. \citealt{Garofalo09,Garofalo10}; however, see \citealt{DeVilliers05,Tchekhovskoy12}). Of course, the possible retrograde spin
obtained with model B depends not only on the assumption that the
potential absorption feature is real, but also that the inner radius
of the accretion disc is coincident with the ISCO of the black hole. It is well accepted that at
low accretion rates the inner disc actually truncates at some
distant radius from the ISCO, although the exact point at which this
begins to happen is still hotly debated \citep[e.g.][]{Reis10lhs}.
The retrograde spin constraint obtained could simply imply instead
that the source was observed when the disc was mildly truncated
($R_{\rm in} \sim 10 R_{\rm G}$).

If the 7.2\,keV absorption feature is real, its most likely
identification is a blueshifted Fe {\sevensize XXVI} absorption line
arising in an outflowing disc wind, although the outflow velocity of
$v_{\rm out} \sim$ 10000\,km s$^{-1}$ implied is substantially
higher than the velocities of $\la$1000\,km s$^{-1}$ typically
observed from winds in black hole binaries. Nevertheless, the energy
of the line is not high enough for an identification with nickel,
and invoking any other iron transition would require an even higher
outflow velocity. Outflow velocities of this order, or even larger,
have frequently been claimed in active galaxies (see \emph{e.g.}
\citealt{Pounds09}, \citealt{Reeves09}, \citealt{Tombesi10}), and
there has more recently been a detection of a similarly fast outflow
in the black hole binary candidate IGR\,J17091-3624
(\citealt{King11}).

The slower disc winds more frequently observed in BHBs are observed
predominantly at high disc inclinations, when we are observing the
source close to an edge-on orientation. However, our modelling would
suggest this is not actually the case for XTE\,J1652-453. In this
context, a relatively weak line with an unusually high outflow
velocity might not be so surprising; most numerical simulations of
equatorial disc winds suggest that both the density and outflow
velocity of such outflows evolve with the inclination at which they
are observed, with the winds appearing more tenuous but with higher
outflow velocities when observed at low inclinations (see
\emph{e.g.} \citealt{Proga02}). An interesting alternative origin
for the large energy shift of this potential feature was recently
proposed by \cite{Gallo11}, who demonstrate that if the absorbing
material originates close to the black hole and orbits with the
accretion disc (as would be expected for a wind launched from the
disc), the same effects that skew the emission from this region can
also result in large shifts in the observed energies of absorption
features. However, in order to produce a narrow feature in this way,
the absorbing material may need to be confined to a relatively
narrow range of radii.

We have also investigated the possibility that the
Comptonising corona could actually have a large scale height above
the plane of the accretion disc (model C). In this scenario we find
that the spin cannot be constrained. The primary constraint on the
black hole spin comes from the extent of the red wing in the iron
emission line, which is directly related to the inner disc radius. Different emissivities result in different iron line
profiles, with higher emissivity indices giving rise to stronger (but not more extended red wings) emission at
lower energies compared to low emissivity indices. The low emissivity
index assumed in this scenario naturally makes it more difficult to constrain the spin given the low signal to noise in the present data.

\section{Conclusion}

We have analysed the simultaneous \emph{XMM-Newton} and \emph{RXTE}
data of the black hole candidate XTE J1652-453. The spectra have
been modelled using a sophisticated model which consists of a
powerlaw continuum and a reflection component. We used the
{\sevensize REFBHB} model, which is appropriate for stellar-mass
black holes, to model the reflected continuum. The convolution model
we use allows for the possibility of retrograde spins. The
possibility of a weak absorption line at $\sim$ 7.2 keV has been
confirmed. Assuming this feature is real, it is likely associated
with a blue-shifted Fe {\sevensize XXVI} line, having an outflow
velocity of $\sim$ 11100 km s$^{-1}$, which is much higher than the
general value seen in stellar-mass black holes, and is similar to
the newly found source IGR J17091-3624 \citep{King11}.

A broad Fe k$\alpha$ line is also present, but our result
indicates that the spin parameter of the source can only be weakly
constrained under certain circumstances. However, if the corona is compact, as is generally assumed, we
find an upper limit of $a^* \sim 0.5$, broadly consistent with the result of
\citet{Hiemstra11}. Longer exposure and dynamic information would be
extremely helpful in breaking the degeneracies between the spin and
inclination, and allow a much tighter constraint in the spin
parameter.

\section*{Acknowledgements}

RCR thanks the Michigan Society of Fellows and NASA. RCR is supported by NASA through the Einstein Fellowship Program,
grant number PF1-120087. DJW acknowledges the financial support provided by STFC, and
ACF thanks the Royal Society. This work was greatly expedited thanks to the help of Jeremy Sanders in optimising the various convolution models.

\bibliographystyle{mn2e_uw}
\bibliography{1652}

\label{lastpage}
\end{document}